\documentclass[aps,twocolumn,groupedaddress]{revtex4}
\usepackage{times}
\usepackage{amsfonts}
\usepackage{amsmath, amsthm, amssymb}
\usepackage{graphicx}
\usepackage{mdframed}

\newcommand{\vk}{{\boldsymbol{k}}} 

\newcommand{\e}[1]{\mathrm{e}^{#1}}

\newcommand{\eg}{\textit{e.g. }}
\newcommand{\etal}{\emph{et al. }}
\def\i{\mathrm{i}}                            
\definecolor{bb}{rgb}{0.8, 0.8, 1.0}

\begin{document}

\title{\flushleft \Large Superconducting Spintronics\\
\flushleft \normalsize Jacob Linder$^{1}$ and Jason W. A. Robinson$^{2}$}

\begin{abstract}
\flushleft{$^1$\textit{Department of Physics, Norwegian University of
Science and Technology, N-7491 Trondheim, Norway}\\
$^2$\textit{Department of Materials Science and Metallurgy, University of
Cambridge, 27 Charles Babbage Road, Cambridge CB3 0FS, UK}}\\
\text{ }\\
\textbf{Traditional studies that combine spintronics and superconductivity have mainly focused on the injection of spin-polarized
quasiparticles into superconducting materials. However, a complete synergy between superconducting and magnetic orders
turns out to be possible through the creation of spin-triplet Cooper pairs, which are generated at carefully engineered
superconductor interfaces with ferromagnetic materials. Currently, there is intense activity focused on identifying materials
combinations that merge superconductivity and spintronics to enhance device functionality and performance. The results look
promising: it has been shown, for example, that superconducting order can greatly enhance central effects in spintronics such
as spin injection and magnetoresistance. Here, we review the experimental and theoretical advances in this field and provide
an outlook for upcoming challenges in superconducting spintronics.}
\end{abstract}
\maketitle

\small

At the interface between materials with radically different properties, new physical phenomena can emerge. A classical example of such an interface is that between a superconductor and a ferromagnet where the opposing electron orders destructively interfere; however, it turns out that under the right conditions at a superconductor-ferromagnet interface both superconductivity and spin-polarization can unite to create a new superconducting state that offers tantalizing possibilities for spin transport in which Joule heating and dissipation are minimized.
 
Spintronics offers the potential for creating circuits in which logic operations controlled by spin currents can be 
performed faster and more energy efficient \cite{zutic_rmp_05} than the charge-based equivalent in semiconductor transistor technologies. Spintronics is one of the most active areas of research and while it offers control of spin and charge at the nanometer scale, it has also found sensory applications in hard disk drive read heads via the giant magnetoresistance effect \cite{baibich_prl_88, binasch_prb_89}. The idea of combining superconductivity with spintronics has historically focused on the net spin-polarization of quasiparticles in superconductors. It is interesting to note that the first  spin transport experiments \cite{meservey_prl_71, meservey_prb_73, meservey_pr_94} involved ferromagnet-superconductor bilayers and pre-dated non-superconducting spin transport experiments \cite{johnson_prl_85}. As will be discussed in this review, it is possible to create pseudo-chargeless spin-1/2 excitations in superconductors \cite{kivelson_prb_90} which have extremely long spin lifetimes.
    
Recently, a more complete synergy between superconductivity and spintronics has been made possible through the discovery of spin-triplet Cooper pairs at superconductor-ferromagnet interfaces. Non-superconducting spin currents are generated by passing charge currents through ferromagnetic materials. As will be explained in this review, spin currents can also be generated by passing supercurrents through ferromagnetic materials. Charge flow within superconductors is carried by Cooper pairs which consist of interacting pairs of electrons \cite{bcs}. The idea of combining superconducting and magnetic order was inititated in the late 1950s when Ginzburg \cite{ginzburg} demonstrated theoretically that the electrons within a Cooper pair in a conventional superconductor will eventually be torn apart due to the so-called orbital effect: in the presence of a magnetic field, the Lorentzian force acts differentially on the oppositely aligned electron spins of a pair. Moreover, the Zeeman interaction between spins and a magnetic field favors a parallel alignment, meaning that for a strong enough magnetic field the pairs are energetically unstable as one electron of a pair is required to spin-flip scatter. However, there exists a way to avoid this problem. The two-fermion correlation function $f$ describing Cooper pairs is subject to the Pauli principle, meaning that the spin-part does not necessarily have to be in a spin-singlet \cite{bcs} antisymmetric state $(\uparrow\downarrow - \downarrow\uparrow)$. So long long as $f$ is antisymmetric under an overall exchange of fermions $1\leftrightarrow 2$, which includes the space, spin, and time coordinates of the two electrons, the Pauli principle is satisfied. This means that Cooper pairs can reside in a spin-triplet state which is symmetric under fermion exchange - that is, $\frac{1}{\sqrt{2}}(\uparrow\downarrow +\downarrow\uparrow)$, $\uparrow\uparrow$, or $\downarrow\downarrow$ - as long as $f$ changes sign under an exchange of space- and time-coordinates as well, allowing for odd-in-time (or odd-frequency) pairing \cite{berizinskii, abrahams_prb_95, coleman_prb_94}. Such a spin-triplet state can coexist with a magnetic field since the Zeeman interaction due to the magnetization is no longer having a pair-breaking effect on the Cooper pairs so long as the orbital effect is suppressed.

Since Cooper pairs can be spin-polarized, it follows therefore that triplet supercurrents can carry a net-spin component and so offer the potential to eliminate the heating effects associated with spintronic devices. However, in order to use such supercurrents in spintronics it is necessary to be able to generate and manipulate triplet pairs in devices. In recent years there has been significant progress in this area, not least on the experimental side where the generation of triplet pairs in superconductor-ferromagnet (SF) structures is becoming routine. 

\begin{figure*}
\includegraphics[width=6.5in]{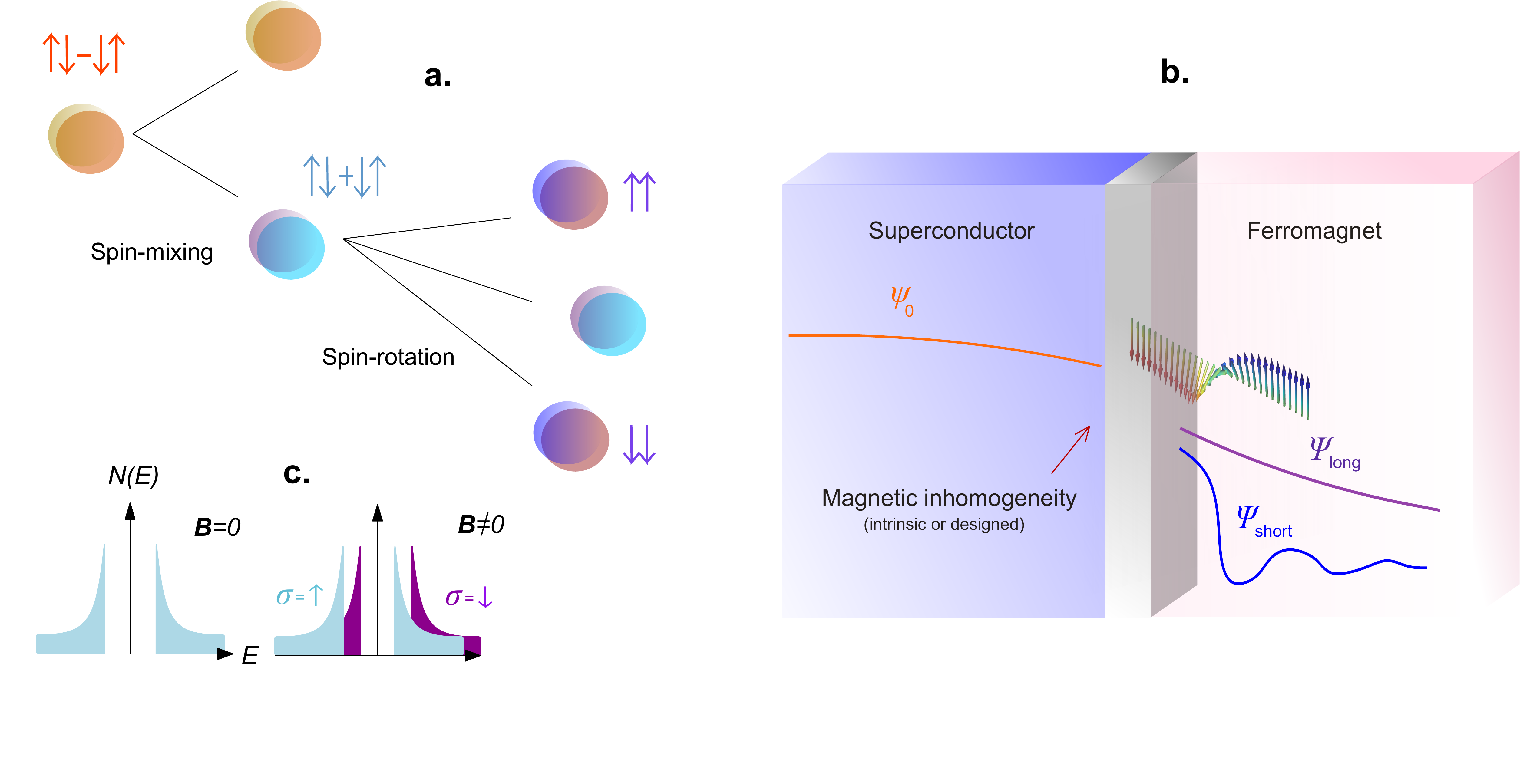}
\caption{\textbf{Figure 1 $\mid$ Cooper pair conversion from a spin-singlet state to a spin-triplet state \&  spin-charge separation in superconductors.} \textbf{a.} Spin-mixing generates spin-zero ($S_z=0$) triplet pair correlations from spin-singlet $S=0$ superconductivity. If spin-rotation occurs due to a change in the quantization axis, $S_z=\pm 1$ triplet pairs form from the $S_z=0$ triplets. \textbf{b.} Starting out with a conventional $s$-wave superconductor which proximity couples a homogeneous ferromagnet, the singlet $\psi_0$ and short-ranged triplet $\Psi_\text{short}$ pair correlations $S_z=0$ rapidly decay in an oscillatory way in the ferromagnet. In the presence of magnetic inhomogeneity at the interface, long-range triplet correlations $\Psi_\text{long}$ emerge in the ferromagnet. \textbf{c.} The relative spin and charge of quasiparticles within a superconductor depends on the energy $E$ of the quasiparticles: near the gap edge, the quasiparticles carry spin but not charge. The density of states $N(E)$ for the spins can be separated by applying in-plane magnetic fields which induces a Zeeman-splitting of the superconducting $N(E)$.}
\label{fig:model}
\end{figure*}

One of the aims of superconducting spintronics involves identifying ways to enhance central effects in spintronics by introducing superconducting materials and to understand the interactions that arise when superconducting and magnetic order coexist. The results look promising: the existence of spin-polarized supercurrents has been verified; spin-polarized quasiparticles injected into superconductors have been shown to have spin-lifetimes that exceed that of spin-polarized quasiparticles in normal metals by several orders of magnitude; and that superconducting spin-valves offer colossal magnetoresistance effects and can switch on and off superconductivity itself. Even magnetization dynamics have been demonstrated to be strongly influenced by superconducting order, raising  the possibility that superconductivity can influence domain wall motion. 

The recent experimental and theoretical advances described above serve as a motivation for the present review. First, we will overview the microscopic mechanisms and theoretical framework which explain how superconducting order and spin-polarization can be reconciled and, secondly, we will discuss a few of the promising proposals which highlight the benefits of superconductivity for spintronics. We also discuss the experimental scene in terms of spin-polarized quasiparticles in superconductors and triplet Cooper pair generation. Finally, we look ahead at promising future directions and outline some of the outstanding issues that need to be addressed in order to develop the field of superconducting spintronics.
\text{ }\\

\noindent\textbf{Spin-flow in superconductors}\\
A key requirement for spintronics is that the spin degree of freedom relaxes slowly enough in order for the spin to be manipulated and read out. Spin lifetimes are nevertheless typically quite short in diffusive materials due to spin-orbit and spin-flip scattering processes which
lead to spin randomization. Another major hurdle relates to the fact that since electrons carry spin and charge, they are susceptible to
processes which cause dissipation and decoherence due to the charge degree of freedom. Finding ways to prolong spin lifetimes in materials is therefore a high priority in spintronics. Superconductors can help resolve this problem. To see why, consider excitations in the superconducting state. Below the energy gap $\Delta$ stable excitations do not exist whereas quasiparticles may be created with energies above the gap. As shown in Box 1, these quasiparticles are always spin-1/2 regardless of their excitation energy, but their effective charge varies strongly with energy $E$. For large energies $E\gg\Delta$, the excitations in 
a superconductor are electron- or hole-like in character. For energies close to the gap edge $E\simeq\Delta$, however, the weight of the electron- and hole-character is almost 
identical. Consequently, they carry a net spin component in the near-absence of charge above the superconducting gap. In addition, their average speed is greatly reduced in the same energy range meaning it takes them longer to scatter through processes involving spin-orbit impurities compared to their scattering rates in the normal state. The net consequence of the above is that the spin lifetime of quasiparticles near the gap edge $E=\Delta$ in a superconductor can be increased by many orders of magnitude compared to within ferromagnetic metals, which is precisely the desirable property sought in spintronics. The realization of spin-charge separation for quasiparticles in superconductors dates back to Kivelson and Rokhsar \cite{kivelson_prb_90} and the spin injection properties in superconducting spin-valve hybrid structures was later studied theoretically in detail by Takahashi \etal \cite{takahashi_prl_99}. Johnson demonstrated the first experimental evidence of non-equilibrium spin injection in the same geometry \cite{johnson_apl_94}. 

\begin{figure*}[t!]
\includegraphics[width=7.0in]{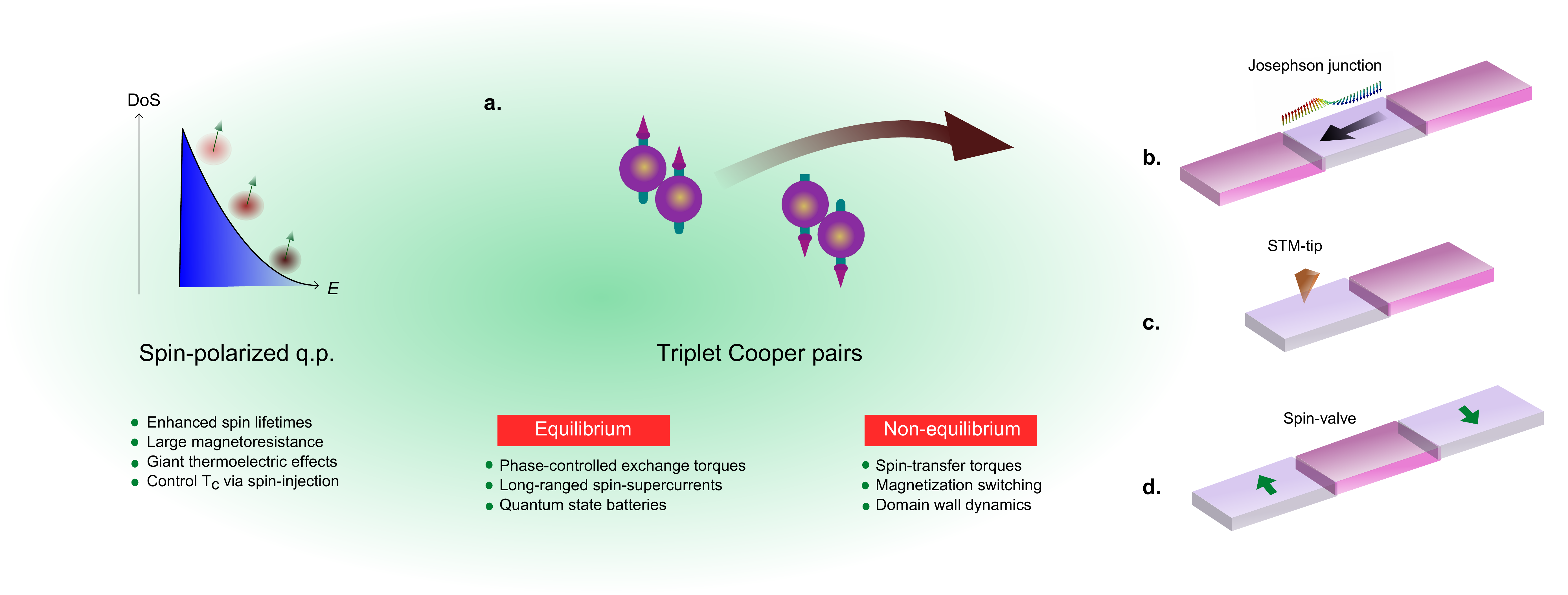}
\caption{\textbf{Figure 2 $\mid$ Applications of superconducting spintronics.} \textbf{a.} Schematic overview of different ways to utilize superconducting spintronics via spin-polarized quasiparticles and triplet Cooper pairs, both in equilibrium and non-equilibrium settings. The fading color of the quasiparticles in the superconducting region represents their loss of effective charge as they approach the gap edge. \textbf{b.-d.} Schematics for typical experimental setups used in superconducting spintronics, including Josephson junctions, bilayers, and spin-valves. }
\label{fig:overview}
\end{figure*}

Theoretical investigations of hybrid structures involving superconductors and ferromagnets were pioneered in the late 1970s by Bulaevskii and Buzdin \cite{bulaevskii}. When a superconductor is placed in good contact with a metal, the tunneling of electrons across the interface results in a proximity effect: the leakage of superconducting pair correlations into the metal and non-superconducting electrons into the superconductor. If the metal is non-magnetic, the pair correlations decay monotonically on the normal metal layer thickness; however, for a ferromagnet the pair correlations decay in an oscillatory manner \cite{buzdin_rmp_05} superimposed on an exponential decay since the Fermi surfaces for spin-$\uparrow$ and spin-$\downarrow$ electrons are no longer degenerate, meaning that the Cooper pairs acquire a finite center-of-mass momentum.

Due to spin-dependent scattering at the interface between the superconducting and ferromagnetic regions, triplet pairing correlations are created (see Figure 1) which decay on a length-scale of the singlet pair correlations (typically a distance of 1-10 nm from the superconductor-ferromagnet interface). Such triplet pairs do not carry any net spin-projection along the quantization axis and so do not appear to have any immediate use in spintronics. In 2001 it was demonstrated in a seminal work  \cite{bergeret_prl_01} (see also Refs. \cite{kadigrobov_epl_01, bergeret_rmp_05}) that triplet pairs that carry spin in addition to charge could also form by introducing magnetic inhomogeneities at the superconductor-ferromagnet interface. The process of converting a spin-singlet Cooper pair into a spin-triplet pair can be understood by introducing the concepts of spin-mixing and spin rotation \cite{eschrig_prl_03} as described in Box 2 and Figure 1. The spin mixing process generates the $S_z=0$ triplet component from a spin-singlet source via spin-dependent phase-shifts that the electrons experience when propagating through a ferromagnetic region or when scattered at a ferromagnetic interface. When the magnetization of the system is textured such that the spin-quantization axis spatially varies, the effect of spin-rotation comes into play thus causing the different spin-triplet components to transform into each other. Through this process spin-polarized Cooper pairs form where both electrons of a pair have the same sign of spin. When propagating through a ferromagnet, the Zeeman field no longer has a pair breaking effect and so triplet Cooper pairs are long-ranged in ferromagnetic materials and have been demonstrated to extend up to hundreds of nanometres even in half-metallic compounds \cite{keizer_nature_06}. The history of long-ranged spin-polarized supercurrents has been covered in detail in Ref. \cite{eschrig_phystoday}. 

There are other ways to generate long-ranged spin-triplet correlations in ferromagnetic structures that are not textured (see examples in Table 1). If a superconducting material lacks an inversion center (either due to its crystal structure or due to the geometry of the setup) it will generally feature antisymmetric spin-orbit coupling such as Rashba spin-orbit coupling \cite{rashba_soc}. This leads to a mixing of excitations from the two spin-bands in such a fashion that spin is no longer a conserved quantity. Instead, the long-lived excitations now belong to pseudospin bands that may be thought of as momentum-dependent combinations of the the original spin species. As a result, the superconducting pairing state in noncentrosymmetric superconductors will intrinsically  be a mixture of singlet and triplet pair correlations \cite{gorkov_prl_01}. When pairing occurs between the quasiparticle excitations of a simple Hamiltonian featuring antisymmetric spin-orbit coupling such as $\hat{H} = \varepsilon_\vk + \boldsymbol{g}_\vk \cdot \boldsymbol{\sigma}$ where $\varepsilon_\vk$ is the normal-state dispersion, $\sigma$ is the Pauli matrix vector, and $\boldsymbol{g}_\vk = -\boldsymbol{g}_{-\vk}$ is a vector characterizing the spin-orbit coupling, the triplet part of the superconducting pairing generally may be described by the relation $\boldsymbol{d}(\boldsymbol{k}) \parallel \boldsymbol{g}(\boldsymbol{k})$ where $\boldsymbol{d}(\vk) \equiv [(\Delta_{\downarrow\downarrow}(\vk)-\Delta_{\uparrow\uparrow}(\vk))/2, -\i(\Delta_{\uparrow\uparrow}(\vk)+\Delta_{\downarrow\downarrow}(\vk))/2,\Delta_{\uparrow\downarrow}(\vk)]$ is the triplet $d$-vector \cite{legget_rmp_75} associated with the spin of the Cooper pair state $\langle \boldsymbol{\sigma} \rangle \propto \i \boldsymbol{d}(\boldsymbol{k})  \times \boldsymbol{d}(\boldsymbol{k}) ^*$.  We emphasize here that the $d$-vector formalism is very suitable to describe also the proximity-induced triplet correlations in superconductor-ferromagnet structures, where the anomalous Green's functions $f_{\sigma\sigma'}$ take on the role of the gaps $\Delta_{\sigma\sigma'}(\vk)$ above. One may thus define a "proximity"' triplet vector $\boldsymbol{f}$. As shown in Ref. \cite{annunziata_prb_12}, the proximity effect between such a system and a homogeneous ferromagnet will thus produce both short-ranged and long-ranged triplet superconductivity inside the ferromagnetic region based on if the spins of the triplet Cooper pairs are perpendicular to or aligned with the Zeeman field. The generation of long-ranged spin-triplets via spin-orbit coupling and homogeneous ferromagnetism has also been expressed in terms of an analogy between D'yakonov-Perel \cite{dp} spin relaxation and precession of spins in normal systems and diffusive systems with antisymmetric spin-orbit coupling in contact with $s$-wave superconductors \cite{bergeret_prb_13}. More specifically, a comparison between the quasiclassical Usadel equation \cite{usadel} (which determines the superconducting pairing correlations quantified by the anomalous Green's function $\boldsymbol{f}$) in the presence of such spin-orbit interactions and the spin diffusion equation for normal state systems (which determines the spin density $\boldsymbol{S}$) shows that the spin-orbit interaction affects the components of $\boldsymbol{f}$ and $\boldsymbol{S}$ in the same way.

We note in passing that using spin-orbit coupling as a source of singlet-triplet mixing has been a central ingredient in proposals related to the emergence of Majorana fermions in condensed matter systems \cite{sau, alicea}.

While the interaction of conventional spin-singlet superconductors
and ferromagnets may result in spin-triplet pairs, they can also be created in bulk spin-triplet superconductors such as Sr$_2$RuO$_4$ \cite{maeno_science_04}
and ferromagnetic superconductors such as the uranium based heavy-fermion compounds \cite{saxena_nature_00, aoki_nature_01} . This includes the creation of
spin currents without resistance \cite{asano_prb_05, gronsleth_prl_06, brydon1, brydon2, tanaka_prb_09} and spin-valve devices controlling the resistance
of the junction via the superconducting critical temperature $T_c$ \cite{romeo_prl_13}. There are, however, practical problems to overcome in order to use triplet superconductors
rather than conventional superconductors for spintronics, such as the requirement of high pressures or sub-Kelvin critical temperatures. Interestingly, the first prototype of a triplet superconductor-ferromagnet bilayer structure (see Figure 2c) was very recently experimentally reported \cite{anwar_arxiv_14}, which may be the first step toward investigating the interface between spintronics and bulk triplet superconductors.
  
\text{ }\\
\noindent\textbf{Spin-polarized quasiparticles and magnetoresistance}\\
\noindent The application of superconducting elements in spintronics necessarily requires non-equilibrium transport driven via \eg voltages or temperature gradients. In this section, we review experimental advances in both equilibrium and non-equilibirum transport and discuss recent theoretical insights which are yet to be realized experimentally.

\begin{figure*}[t!]
\includegraphics[width=7.0in]{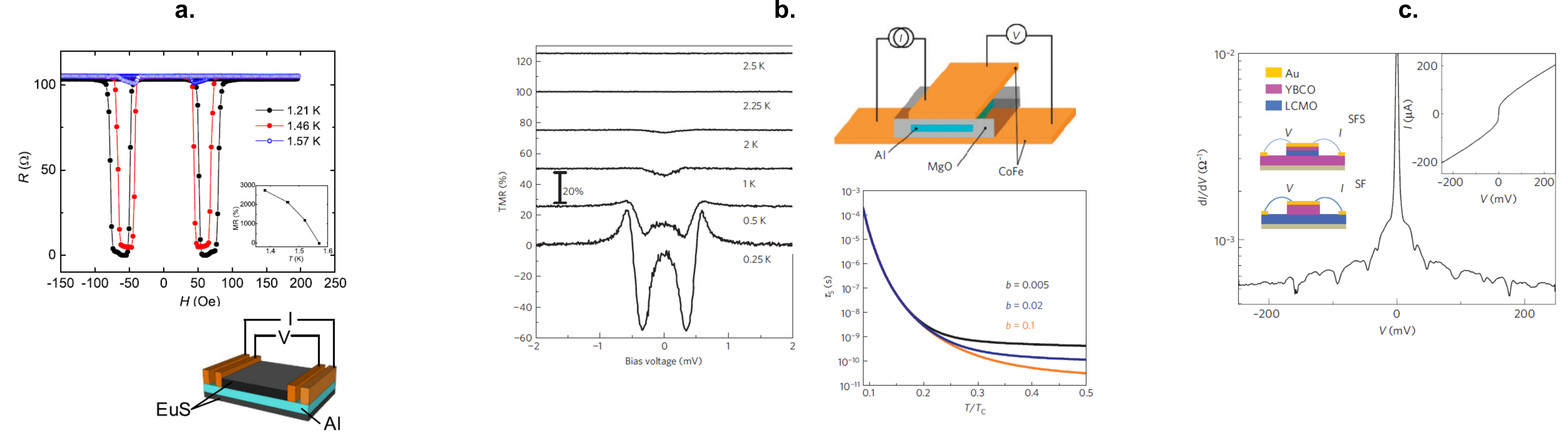}
\caption{\textbf{Figure 3 $\mid$ Recent experimental highlights for superconducting spintronics.} \textbf{a.} Infinite magnetoresistance effect in a superconducting spin-valve with ferromagnet insulators. Reprinted and adapted figure with permission from Li \textit{et al.}, Phys. Rev. Lett. \textbf{110}, 097001 (2013) \cite{li_prl_13}. Copyright (2013) by the American Physical Society. \textbf{b.} Evidence of an extremely large spin lifetime as probed via tunnel magnetoresistance oscillations due to spin imbalance in the superconducting state. Adapted from Yang \textit{et al.}, Nature Materials \textbf{9}, 586 (2010) \cite{yang_nature_10}. \textbf{c.} Spectroscopic signature of long-ranged triplet correlations in a half-metal with quasiparticle interference giving rise to conductance-oscillations. Adapted from Visani \textit{et al.}, Nature Physics \textbf{8}, 539 (2012) \cite{visani_nphys_12}.}
\label{fig:overview}
\end{figure*}

We begin by discussing effects related to spin-polarized quasiparticles in superconductors. Although early studies of spin imbalance in superconducting spin-valves assumed that the spin-lifetime in the superconducting state $\tau_s$ was unchanged \cite{chen_prl_02} from the normal state $\tau_n$, more recent experiments have demonstrated greatly enhanced quasiparticle spin-lifetimes  in the superconducting state. For example, Yang \etal \cite{yang_nature_10} reported
spin lifetimes of a non-equilibrium spin density in superconducting Al that were a million times longer than in the normal state by measuring a considerable tunnel magnetoresistance due to spin imbalance that could only be consistent with a very large spin lifetime. The spin-charge separation and reduced spin-orbit scattering rate
near the gap edge for quasiparticles in a superconductor leads to strongly increased spin lifetimes compared to the normal state due to their movement slowing down greatly at this energy range (see Box 1 for discussion). Importantly, the enhancement of the spin density lifetime in the superconducting state relative the normal state becomes much larger when accounting for impurity spin-orbit scattering \cite{yang_nature_10} in the relative spin susceptibility $\chi_S/\chi_N$, which in this case remains finite as $T\to 0$ (see Figure \ref{fig:overview}b). A treatment without spin-orbit effects, on the other hand, provides a much smaller increase of the spin-lifetime in the superconducting state relative the normal state \cite{yamashita_prb_02}. Using a slightly different setup where an intrinsic Zeeman-splitting was induced in the superconducting region via in-plane magnetic fields,  Quay \etal \cite{quay_nphys_13} showed evidence of a nearly chargeless spin imbalance in superconducting Al using a spin-valve setup with Co ferromagnets. Their measurements of the non-local resistance due to diffusion of the spin imbalance revealed vastly different timescales for spin and charge
relaxation of 25 ns versus 3 ps. In addition, their results implied a strongly enhanced spin lifetime in the superconducting state, $\tau_s \simeq 500 \tau_n$. The intrinsic spin-splitting of the density of states permitted a strong spin accumulation of fully polarized spins when the tunneling from an F electrode matched the gap edge for one of the spin species. Similar conclusions were also reached by H{\"u}bler \etal \cite{hubler}.

It is important to note that the change in spin-relaxation length $\lambda_\text{sf}$ in the superconducting state compared to the normal state depends on the origin of the spin-flip processes. For spin-orbit scattering via impurities, $\lambda_\text{sf}$ is predicted to be the same both above and below $T_c$ \cite{yamashita_prb_02} although Poli \etal \cite{poli_prl_08} reported a decrease of $\lambda_\text{sf}$ by an order of magnitude in the superconducting state which was attributed to spin-flip scattering from magnetic impurities \cite{morten_prb_04}. Information about the spin-relaxation length was obtained by non-local resistance measurements that could probe the diffusion of the spin imbalance that originated at the spin injection point. We also note that spin absorption by superconductors with strong spin-orbit coupling has very recently been demonstrated by Wakamura \etal \cite{wakamura_prl_14}, where the spin relaxation time was found to be much greater in the superconducting state of Nb compared to its normal state.\\

Another example of how superconducting order can enhance conventional spintronics is through the magnetoresistance effect. In the superconducting analog of a spin-valve device, the metallic spacer between two ferromagnets is replaced with a superconductor. The magnetization configuration influences the resistance experienced by an injected current just as it does in non-superconducting device, but here it can also switch on and off the superconducting state which corresponds to an infinite magnetoresistance. The earliest theory investigation of a superconducting spin-valve setup dates back to de Gennes \cite{pdg} whereas experiments \cite{hauser} soon after confirmed his prediction of a higher $T_c$ in the anti-parallell state of the ferromagnets compared to the parallell configuration. When the superconductor is sufficiently thin, a proximitised ferromagnet will influence the superconducting state in the following way. \\
\text{ }\\
 
\noindent\fcolorbox{black}{bb}{
\begin{minipage}[l]{0.97\linewidth}
\small
\textbf{Box 1 $\mid$ Spin injection and spin imbalance in superconductors}. The quasiparticle excitations in a superconductor can be described by $4\times1$ spinors when considering both particle-hole and spin space. The excitations are in general mixture of electron- and hole-states, carrying a weight from each of these branches in their wavefunction. Nevertheless, they are typically characterized as being electron- or hole-like depending on the asymptotic behavior of the wavefunction for energies $E\gg\Delta$. For instance, an electron-like quasiparticle with spin-$\uparrow$ may be written as $\psi = [u,0,0,v]^\text{T}\e{\i q_e x},$where $u (v) = \sqrt{\frac{1}{2}(1 + (-) \sqrt{E^2-\Delta^2}/E)}$. For $E\gg\Delta$, $u\to1$ and $v\to0$. The wavevector of the excitation is $q_e = \sqrt{2m(\mu + \sqrt{E^2-\Delta^2})}$ for a simple parabolical normal-state dispersion relation $\varepsilon_k = k^2/2m^*$ where $m^*$ is an effective mass. The spin and charge content of this quasiparticle can be evaluated by introducing the operators 
\begin{align}
\hat{\mathbf{S}} = \frac{\hbar}{2}\begin{pmatrix}
\boldsymbol{\sigma} & \boldsymbol{0}\\
\boldsymbol{0} & -\boldsymbol{\sigma}^* \\
\end{pmatrix},\; \hat{Q} = -|e|\begin{pmatrix}
\boldsymbol{1} & \boldsymbol{0}\\
\boldsymbol{0} & -\boldsymbol{1}\\
\end{pmatrix},
\end{align}
where $|e|$ is the magnitude of the electron charge and $\boldsymbol{\sigma}$ is a vector with the Pauli spin matrices as components. Computing the expectation values for spin and charge using $\psi$ above then yields:
\begin{align}
\langle \hat{\mathbf{S}} \rangle  = (\hbar/2)\hat{\mathbf{z}},\; \langle \hat{Q} \rangle = -|e|\sqrt{E^2-\Delta^2}/E.
\end{align}
It is seen that while the spin of quasiparticles is constant, the effective charge is strongly dependent on its excitation energy $E$ and vanishes near the gap edge $E\to\Delta$. This is the key property of the excitations which cause spin-charge separation and  enhanced spin lifetimes within superconductors. The group velocity $v_g=\frac{\partial E}{\partial k} = \frac{k}{m^*} \frac{\varepsilon_k-\mu}{E}$ of the excitation $E = \sqrt{(\varepsilon_k-\mu)^2+\Delta^2}$ is also very small near the gap edge since $E\to\Delta$ implies $(\varepsilon_k-\mu) \to 0$, causing scattering events to be less frequent and thus the lifetime to increase.  With regard to spin current injection into a superconducting spin-valve (see Figure 2d), the resulting spin imbalance in the superconductor depends strongly on the magnetization configuration. Following Ref. \cite{takahashi_prl_99}, for a superconductor of smaller thickness than the spin diffusion length, the spin-$\uparrow$ and spin-$\downarrow$ distribution functions for quasiparticles can be taken as spatially uniform and described by the Fermi-Dirac function $f(E)$, but with shifted chemical potentials. In the P alignment, the spin conductances $G_\sigma$ are equal at both interfaces due to the symmetric setup and there is no net shift $\delta\mu$ in the chemical potential for any of the spin species $\sigma$. For the AP alignment, the different density of states for spin-$\uparrow$ and spin-$\downarrow$ at the two interfaces gives rise to imbalanced spin currents and produces a net shift in chemical potential for spins $\sigma$ inside the superconductor. One may write $f_\uparrow(E) = f_0(E-\delta\mu)$ and $f_\downarrow(E) = f_0(E+\delta\mu)$. Upon evaluating the self-consistency equation for the superconducting order parameter, $1 = gN_0 \int^{\omega_D}_0 \text{d}\varepsilon E^{-1} (1-f_\uparrow-f_\downarrow)$, it is seen that the spin-discriminating shift in chemical potential takes an equivalent role of a Zeeman splitting $\mu_B H$ due to an external field $H$, causing a first-order phase transition at the Clogston-Chandrasekhar \cite{cc1,cc2} limit $\mu_B H = \Delta_0/\sqrt{2}$. Above, $\varepsilon$ is the normal-state dispersion, $g$ is the attractive pairing potentialy, $N_0$ is the normal-state DoS at the Fermi level, $\mu_B$ is the Bohr magneton, while $\omega_D$ is the Debye cut-off.
\end{minipage}}

\newpage

Even in the absence of a potential gradient, the superconducting critical temperature $T_c$ is non-monotonic and, in certain cases, reentrant on ferromagnetic layer thickness $d_F$ \cite{tagirov_prl_99,jiang_prl_95, muhge_prl_96, oh_apl_97}. The strong oscillatory dependence of $T_c$ on $d_F$ may be understood in terms of quasiparticle interference inside the ferromagnetic region \cite{fominov_prb_02}. This effect is most pronounced when the superconductor thickness $d_S$ creeps below the superconducting coherence length $\xi_S$, suggesting that the inverse proximity effect (the induction of ferromagnetic order inside the superconductor) is responsible for this phenomenon.

A variation of $T_c$ on $d_F$ requires the measurement of multiple samples but controlling $T_c$ through the relative orientation of the F layers in an FSF spin-valve can be achieved within a single device \cite{zhu_prl_10, leksin_prl_12, jara_prb_14, banerjee_nature_14}. Generally one expects the AP configuration of the F layers to be more compatible with spin-singlet pairing than the P configuration: when the thickness of the S layer is comparable to the superconducting coherence length $\xi_S$, the electrons in a singlet pair feel a reduced Zeeman field but in the P state the fields are additive and so $T_c$ is suppressed as confirmed by Gu \etal \cite{gu_prl_02} and Moraru \etal  \cite{moraru_prl_06}. When the magnetizations are non-collinear the $T_c$ behaves non-monotonically on the angle between the F layers, displaying a minimum at a relative misalignment angle of $\pi/2$ \cite{leksin_prl_12, jara_prb_14, wang_prb_14} due to the generation of triplet pairs. 
Such an effect can be understood qualitatively from the fact that the proximity-induced triplet pairing was theoretically found to be 'anti-correlated' to the change in $T_c$ \cite{jara_prb_14}: with more singlet Cooper pairs leaking into the ferromagnetic side (suppression of $T_c$), triplet pairing becomes enhanced. Recently, an unusually large change in $T_c$ of order 1 K was reported by using half-metallic ferromagnets in a spin-valve setup \cite{singh_arxiv_14}. In Ref. \cite{li_prl_13}, ferromagnetic insulators were used in contrast to metallic ferromagnets: using an EuS/Al/EuS setup with layer thicknesses of a few nm, a full transition from a superconducting to resistive state (governed by the proximity-induced Zeeman field in the superconductor) was observed upon going from an AP to P configuration, resulting in an infinite magnetoresistance (see Figure \ref{fig:overview}a). Large changes in $T_c$ have also been reported for V/Fe spin-valves \cite{westerholt_prl_05, miao_prl_08}.
  
The control of $T_c$ of superconducting spin-valves is generally achieved without applying an intentional voltage bias and is therefore due to the proximity effect. In non-equilibrium situations where voltages are applied, spin injection or transport measurements can be performed to assess how superconductivity modifies spin transport. Several experiments have considered a superconducting spin-valve setup in which a bias voltage is applied between metallic ferromagnets \cite{gu_prl_02, potenza_prb_05, moraru_prl_06, pena_prl_05, yang_nature_10}. In the presence of tunneling barriers which suppress the proximity effect, the role of the magnetization configuration can be reversed compared to the case when no voltage is applied. In the P state, the injected spin from one ferromagnet provides the output in the second ferromagnet and no net spin imbalance occurs in the superconducting region. The superconducting gap is thus unaffected by the spin injection irrespective of the bias voltage applied. This changes in the AP state: due to the different density of states for the majority and minority spins in the two ferromagnetic regions, spin injection from one ferromagnet cannot be compensated by an outflow of spin in the other which results in a net spin imbalance in the superconductor. The superconducting state is therefore weakened is and ultimately destroyed by increasing the voltage $V$ \cite{takahashi_prl_99}. The spin imbalance can in turn be detected via magnetoresistance measurements. 

\text{ }\\
\noindent\textbf{Triplet Cooper pairs and magnetization dynamics}\\
An interesting prospect that emerges from the combination of magnetic and superconducting order is that of spin-supercurrents. If Cooper pairs are spin-polarized they should be able to transport not only charge, but also a net spin component but without dissipation.  A number of proposals have been put forward to explain how spin-supercurrents can be created and controlled in hybrid structures, including Josephson junctions (see Figure 2b) with domain walls or textured ferromagnets \cite{eremin_prb_06, halasz_prb_11}, bilayer and trilayer ferromagnetic regions \cite{trifunovic_prb_11}, spin injection \cite{malshukov_prb_12}, and via spin-active interfaces \cite{eschrig_nphys_08} where a net interface magnetic moment is misaligned to the bulk magnetization. The first experimental demonstration of long-ranged supercurrents was reported by Keizer \etal \cite{keizer_nature_06} via the observation of supercurrents through the half-metallic ferromagnet CrO$_2$. Since spin-singlet superconductivity cannot penetrate a fully spin-polarized material,  this result necessarily implied the supercurrents were fully spin-polarised. The results were later repeated by Anwar \etal \cite{anwar_prb_10}. In 2010, a series of experiments by different groups  demonstrated systematic evidence of spin-triplet pairing in SFS Josephson junctions: Khaire \etal \cite{khaire_prl_10} used ferromagnetic/non-magnetic multilayer spin-mixers which were positioned at both superconductor interfaces while Robinson \etal \cite{robinson_science_10} used the helical rare earth antiferromagnet Ho in order to generate triplet supercurrents in Co, and Sprungmann \etal \cite{sprungmann_prb_10} utilized a Heusler alloy in order to generate triplet supercurrents. All of these experiments share similarities to the SF'FF'S device proposed by Houzet and Buzdin \cite{houzet_prb_07} where the F'/F interfaces are magnetically coupled non-parallel.

\text{ }\\

\noindent\fcolorbox{black}{bb}{
\begin{minipage}[l]{0.97\linewidth}
\textbf{Table 1 $\mid$ Emergent superconducting correlations in generic hybrid structures.} Consider a $S/X/Y$ structure  where $S$ is an $s$-wave superconductor, $X$ is the layer separating the two materials and $Y$ is a material with certain properties as tabulated below. We allow for $X$ to be an insulator which is either non-magnetic or spin-polarized with a misaligned moment compared to the magnetization in $Y$, denoting the latter as spin-active. F stands for ferromagnet, SOC for antisymmetric spin-orbit coupling (such as Rashba type), while $\psi_0$ denotes spin-singlet Cooper pairs while $\Psi_\text{short/long}$ denotes short-ranged and long-ranged triplet Cooper pairs.\\
\text{ }\\
\begin{tabular}{ l| c | c}
   \hline       
   \textbf{Material $Y$} & \textbf{Insulating $X$} & \textbf{Spin-active $X$} \\
	 \hline
   Normal metal & $\psi_0$ & $\psi_0$ + $\Psi_\text{long}$\\
   Homogeneous F & $\psi_0$ + $\Psi_\text{short}$ & \;$\psi_0$ + $\Psi_\text{short}$ + $\Psi_\text{long}$ \;\\
	 Homogen. F + SOC & \; $\psi_0$ + $\Psi_\text{short}$ + $\Psi_\text{long}$ \; & \; $\psi_0$ + $\Psi_\text{short}$ + $\Psi_\text{long}$ \; \\
	 Inhomogeneous F & \; $\psi_0$ + $\Psi_\text{short}$ + $\Psi_\text{long}$ \; & \; $\psi_0$ + $\Psi_\text{short}$ + $\Psi_\text{long}$ \; \\
	 Half-metallic F& None & $\Psi_\text{long}$\\
 \end{tabular}
\end{minipage}}

\text{ }\\

Although it is now established that triplet supercurrents exist, their most interesting property - spin - is only inferred indirectly from supercurrent measurements. In conventional spintronics, it is known that spin-currents cause effects such as spin-transfer torque switching of magnetic elements and magnetization dynamics and so the observation of similar effects due to triplet supercurrents would confirm the net spin of triplet pairs and would therefore pave the way for applications. Several theoretical works have considered such situations and demonstrated that triplet supercurrents can indeed induce spin-transfer torque switching \cite{waintal_prb_02, zhao_prb_08} and magnetization dynamics in the superconducting state \cite{buzdin_prl_09, linder_prb_11, teber_prb_10, holmqvist_prb_11, kulagina_prb_14}. Furthermore, the influence of superconductivity on spin-pumping effects have been theoretically investigated both in Josephson junctions \cite{houzet_prl_07}  and in SF bilayers \cite{yokoyama_prb_09}. Other works have discussed spin dynamics in Josephson junctions \cite{zhu_prl_04} and the possibility of using spin-polarized supercurrents to induce magnetic domain wall motion \cite{sacramento1, sacramento2, linder_prx_14}. Magnetic domain wall motion is a major research theme in spintronics as it can offer an alternative way to transmit and store information in a non-volatile way. It has been shown in Ref. \cite{linder_prx_14} that domain wall motion in superconducting junction can control whether the system resides in a dissipative or lossless state by locally switching on or off the superconductivity. The enhancement of supercurrents through the generation of triplet Cooper pairs when passing through a magnetic domain wall was experimentally demonstrated in Ref. \cite{robinson_nature_12}. Another work \cite{baker_arxiv_14} proposed to make use of exchange spring magnetic systems where the magnetization texture is tunable via an external field which in turn triggers transitions between 0 and $\pi$ states. The study of superconducting magnetization dynamics is at an early stage, expecially from the experimental side and so there remains much work to be done in this particular area of superconducting spintronics. We note that the current densities required to obtain magnetization switching and domain wall motion in non-superconducting systems can in some cases be achieved with densities as low as 10$^5$ A/cm$^2$, which is comparable with critical current densities reported in superconductor-ferromagnet-superconductor junctions. It is clear that domain wall motion would necessitate a non-equilibrium supercurrent setup.

The relation between triplet supercurrents and the spin-transfer torque that they can induce is intricate as they will have a feedback effect on each other \cite{blanter_prl_07}. This was explained by Waintal and Brouwer \cite{waintal_prb_02}: let $F$ be the free energy of a Josephson junction containing two ferromagnetic  layers with magnetization vectors that are misaligned with an angle $\theta$. Denoting the superconducting phase difference as $\phi$, the equilibrium charge- and spin-currents $I_Q$ and $I_S$ at a finite temperature are given by $I_Q = \frac{2e}{\hbar}\frac{\partial F}{\partial \phi}$ and $I_S = \frac{\partial F}{\partial \theta}.$ Note that the equilibrium spin current is formally equivalent to a torque $\tau$ acting on the magnetizations which is equal in magnitude but opposite in sign for the two layers. Upon combining these equations, one finds that
\begin{align}
\frac{\partial I_Q}{\partial \theta} = \frac{2e}{\hbar} \frac{\partial \tau}{\partial \phi}.
\end{align}
Since the charge supercurrent depends sensitively on the relative angle $\theta$ between the magnetizations \cite{bergeret_prl_01,  kulic_prb_01, pajovic_prb_06}, the above equation shows that spin-transfer torque is tunable via the superconducting phase difference $\phi$. 

\text{ }\\
\noindent\textbf{Phase-batteries and thermoelectric effects}\\
The combination of superconducting and magnetic order in hybrid structures also produces quantum effects which may find applications in cryogenic spintronics in the form of so called phase battery junctions or$\varphi$-junctions. In a Josephson junction without any magnetic elements the equilibrium phase difference between the superconductors is zero. Introducing a ferromagnet as the interlayer separating the superconductors, opens the possibility of $\pi$-coupling in the equilibrium state as first predicted in \cite{bulaevskii} and experimentally verified in Ref. \cite{ryazanov_prl_01}. However, the quantum ground state phase-difference $\varphi$ between two conventional $s$-wave superconductors separated by a magnetic interlayer is not necessarily 0 or $\pi$, but $0\leq \varphi \leq \pi$. Such a state can consist of either an extra phase shift in the first harmonic of the current-phase relation, providing a non-degenerate minimum for the ground-state \cite{asano_prb_06, margaris, buzdin_prl_08} or doubly degenerate minima $\pm \varphi$ for the ground-state resulting from an interplay between the sign and magnitude of the first two harmonics \cite{buzdin_prb_05, sickinger_prl_12}. The merit of creating a $\varphi$-junction where the equilibrium phase difference is tunable is that it may serve as a phase battery: a device which provides a constant phase shift between the two superconductors in a quantum circuit. Such a junction then supplies a phase shift $\varphi$ in analogy to how a voltage is supplied by a battery, with the important difference that the phase does not discharge since the superconducting currents flowing in the system are dissipationless. In junctions that effectively feature three ferromagnetic layers with misaligned magnetizations, the spin chirality $\chi$ has been demonstrated \cite{asano_prb_06, margaris} to be intimately linked with the realization of a $\varphi$ state: $\chi \equiv \boldsymbol{M}_1 \cdot (\boldsymbol{M}_2 \times \boldsymbol{M}_3).$ However, the $\varphi$-junction may also be realized in other geometries and with homogeneous Zeeman fields in the presence of spin-orbit coupling, as predicted in Ref. \cite{buzdin_prl_08}.  Another example is a magnetic Josephson junction where the interlayer consists of two magnetic regions with different thicknesses and generates a spontaneous fractional vortex state, resulting in a degenerate $\varphi$-state as shown in Ref. \cite{sickinger_prl_12}.  \\
\text{ }\\

Finally, we briefly discuss thermal biasing - thermoelectric - devices for superconducting spintronics. Thermoelectric effects chiefly arise due to the breaking of electron-hole symmetry, a feature most apparent in semiconducting materials where the chemical potential is electrically tunable. In superconductors, electron-hole symmetry is preserved near the Fermi level, and so thermoelectric effects are negligible. However, it is possible to break electron-hole symmetry per spin species $\sigma$ while maintaining the overall electron-hole symmetry by using ferromagnet-superconductor hybrid structures \cite{machon_prl_13, ozaeta_prl_14}, which can lead to large thermopowers and figures of merit. In the presence of spin-selective tunneling, as may be achieved by tunneling to a ferromagnetic electrode, one may also achieve large thermoelectric effects since electron-hole symmetry is  broken for each spin species \cite{kawabata_apl_13, giazotto_apl_14}. \\
\text{ }\\

\noindent\textbf{Outlook and perspectives}\\
\noindent We end the review by offering our perspective on possible directions that may be fruitful to explore in order to develop superconducting spintronics. While progress has been most pronounced on the theoretical understanding of SF proximity effects over the last decade, the experimental activity has in the last few years started to catch up. Nevertheless, there remains a plethora of interesting physics to investigate and we speculate that the most valuable experiments in the near future will directly verify (and quantify) the spin-polarization of triplet states generated by different SF systems - existing experiments provide  compelling evidence for spin-triplet pairing in such structures, but they are not directly probing or using the spin carried by triplet supercurrents. Experiments which, therefore, demonstrate effects such as magnetization switching, magnetization precession, spin-transfer torque, or domain wall motion due to spin-polarized supercurrents will be pivotal in establishing applications of superconducting spintronics. Another issue that deserves investigation is the injection of spin-triplet pairs into superconductors, akin to the injection of spin-polarized quasiparticles into superconductors. Here, tunneling experiments will be essential in order to understand how the density of states in a superconductor is modifed due to the formation of a triplet state - in effect, the inverse of what is usually studied. We also mention that it might be interesting to design more comprehensive theories for the treatment of the ferromagnetic order in superconducting proximity structures, which is usually simply modelled by a Zeeman field $h$ acting on the spins of the electrons. This could be done by incorporating the effect of spin-bandwidth asymmetry (spin-dependent carrier masses) and also by considering more seriously the role of the magnetic vector potential in the proximity effect. We also note that the electromagnetic effect of stray fields in SF structures have been experimentally shown to offer an interesting way to control superconductivity \cite{moshchalkov1, moshchalkov2, villegas_sst_11}.
 
There is also a need to develop spin-triplet theory in order to understand better the interactions between superconducting and spin-polarized order, particularly in non-equilibrium devices where spin and charge dynamics are important. The mechanism required for generating triplet pairs at SF interfaces are generally well understood, just as equilibrium proximity effects are in Josephson junctions and SF multilayers, but in order to advance superconducting spintronics it is essential to develop a framwork for non-equilibrium transport which can account for dynamic interactions involving spin-triplet pairs and ferromagnetic layers \cite{teber_prb_10, holmqvist_prb_11}. Related to this, it is also necessary to clarify the mutual dependence between supercurrent flow and magnetization configuration. The formation of so-called Andreev bound states \cite{andreev} in textured magnetic Josephson junctions should influence the spin-pattern in the ground-state of such systems as they contribute to the effective field $\boldsymbol{H}_\text{eff}$ which in turn determines the equilibrium magnetization profile via the condition $\boldsymbol{m} \times \boldsymbol{H}_\text{eff} = 0$. Whereas the magnetic profile of a junction is usually considered as being fixed, the Andreev bound state contribution is phase-sensitive which suggests that the magnetization texture could be controlled via the superconducting phase difference \cite{kulagina_prb_14}. Moreover, the sizable thermoelectric effects in superconducting hybrids are of practical interest due to the possibility to transform excess heat to electric energy in a highly efficient manner, suggesting applications within cooling of nanoscale systems and thermal sensors/detectors.\\
\text{ }\\
\noindent\fcolorbox{black}{bb}{
\begin{minipage}[l]{0.97\linewidth}
\small
\textbf{Box 2 $\mid$ Spin-mixing and spin-rotation at superconducting interfaces.} The process of generating spin-triplet superconductivity starting out from a spin-singlet Cooper pair can be understood conveniently by drawing upon the phenomena of spin-mixing and spin-rotation \cite{eschrig_prl_03}. The wavefunction for a spin-singlet Cooper pair can be written as 
\begin{align}
\psi_0 = \sqrt{\frac{1}{2}}(\mid\uparrow,\vk\rangle\mid\downarrow,{-\vk}\rangle - \mid\downarrow,\vk\rangle \mid\uparrow,{-\vk}\rangle ),
\end{align}
where the prefactor ensures proper normalization. When the electrons of a Cooper pair encounter an interface region to a ferromagnetic material, scattering at the interface is accompanied not only by a shift in momentum but also a spin-dependent shift $\theta_\sigma$, $\sigma=\uparrow,\downarrow$ in the phase of the wavefunction due to the Zeeman field that splits the majority and minority spin carriers. This may be written as 
\begin{align}
\mid\uparrow,{\vk}\rangle  \to \e{\i\theta_\uparrow}\mid\uparrow,{-\vk}\rangle ,\; \mid\downarrow,{\vk}\rangle  \to \e{\i\theta_\downarrow}\mid\downarrow,{-\vk}\rangle .
\end{align}
Applying these transformations to $\psi_0$ results in a wavefunction which is a superposition of a spin-singlet and $S_z=0$ spin-triplet wavefunction $\Psi_\text{short} \equiv (\mid\uparrow,\vk\rangle\mid\downarrow,{-\vk}\rangle$ + $\mid\downarrow,\vk\rangle\mid\uparrow,{-\vk}\rangle)/\sqrt{2}$. The singlet and triplet parts are weighted by $\cos\Delta\theta$ and $\sin\Delta\theta$ respectively, where $\Delta\theta\equiv\theta_\uparrow-\theta_\downarrow$. In the absence of spin-dependent phase-shifts ($\Delta\theta=0$), the triplet component vanishes. The next step is to generate the equal-spin triplet components $S_z=\pm1$ which are insensitive to the paramagnetic pair-breaking effect of a Zeeman field as the spins of the electrons in the Cooper pair are already aligned. The appeareance of such long-ranged triplet correlations $\Psi_\text{long} \equiv \mid\uparrow,\vk\rangle \mid\uparrow,{-\vk}\rangle $ (or $\mid\downarrow,\vk\rangle \mid\downarrow,{-\vk}\rangle $) can only be brought about by rotating/flipping one of the spins in the $S_z=0$ triplet component. In this sense, the singlet Cooper pairs have served their purpose in terms of generating long-ranged triplets once the short-ranged triplets $\Psi_\text{short}$ have been created and are no longer needed. A magnetic texture serves as a source for spin-rotation which can be seen by letting the quantization axis be aligned with the local magnetization direction. Consider an $S_z=0$ triplet state in a part of the system where the magnetization (and thus quantization axis) points along the $z$-direction. In another part of the system where the magnetization points in the $x$-direction, the same triplet state now looks like a combination of the equal-spin pairing states $S_z=\pm1$ as seen from the new quantization axis. The combination of spin-mixing
and spin-rotation processes then explain how the spin-singlet $s$-wave
component of the bulk superconductor may be converted
into a long-range spin-triplet component that is able to survive
even in extreme environments such as half-metallic ferromagnets that are fully spin-polarized.
\end{minipage}}
\text{ }\\

In summary, we have provided an overview into the past and present activity related to superconducting spintronics, including the associated quantum effects that appear. With advances in experimental fabrication processes and better control of interface properties, there is good reason to be optimistic about further discoveries of novel physics that arise due to the synergy between superconductivity and spintronics.

\text{ }\\
\noindent  \textbf{Acknowledgments}\\
\begin{scriptsize}
\noindent The authors acknowledge useful discussions with J. Aarts, M. Alidoust, M. Aprili, A. Balatsky, W. Belzig, F. Bergeret, A. Black-Schaffer, M. Blamire, A. Brataas, A. Byzdin, L. Cohen, M. Cuoco, M. Eschrig, F. Giazotto, G. Halasz, K. Halterman, I. Kulagina, O. Millo, J. Moodera, N. Nagaosa, E. Scheer, A. Sudb{\o}, Y. Tanaka, T. Yokoyama. J.L. was supported by the Research Council of Norway, Grants No. 205591 and 216700. J.W.A.R. was supported by the UK Royal Society and the Leverhulme Trust through an International Network Grant (IN-2013-033).
\end{scriptsize}
\par
\text{ }\\
\noindent  \textbf{Author contributions}\\
\begin{scriptsize}
\noindent J.L. and J.W.A.R. co-wrote the paper and contributed to all its aspects.
\end{scriptsize}
\par
\text{ }\\
\noindent  \textbf{Additional information}\\
\begin{scriptsize}
\noindent Correspondence and requests for materials should be addressed to J.L. or J.W.A.R.
\end{scriptsize}
\par
\text{ }\\
\noindent  \textbf{Competing financial interests}\\
\begin{scriptsize}
\noindent The authors declare no competing financial interests.
\end{scriptsize}


\begin{thebibliography}{99}
{\scriptsize 

\bibitem{zutic_rmp_05} Zutic, I., Fabian, J., and Das Sarma, S. Spintronics: Fundamentals and applications. \textit{Rev. Mod. Phys.} \textbf{76}, 323 (2004).

\bibitem{baibich_prl_88} Baibich, M. N., Broto, J. M., Fert. A, Nguyen Van Dau, F., Petroff, F., Eitenne, P., Creuzet, G., Friederich, A. and Chazelas, J. Giant magnetoresistance of (001)Fe/(001)Cr
magnetic superlattices. \textit{Phys. Rev. Lett.} \textbf{61}, 2472 (1988).

\bibitem{binasch_prb_89} Binasch, G., Gr{\"u}nberg, P., Saurenbach, F., and Zinn, W. Enhanced magnetoresistance in layered magnetic structures
with antiferromagnetic interlayer exchange. \textit{Phys. Rev. B} \textbf{39}, 4828 (1989).

\bibitem{meservey_prl_71} Meservey, R. \& Tedrow, P. M. Spin-Dependent Tunneling into Ferromagnetic Nickel. \textit{Phys. Rev.  Lett.} \textbf{26},  192  (1971).

\bibitem{meservey_prb_73} Meservey, R. \& Tedrow, P. M. Spin Polarization of Electrons Tunneling from Films of Fe, Co, Ni, and Gd. \textit{Phys. Rev. B} \textbf{7}, 318 (1973)

\bibitem{meservey_pr_94}  Meservey, R. \& Tedrow, P. M. Spin-polarized electron tunneling. \textit{Phys. Rep.} \textbf{238}, 173 (1994).

\bibitem{kivelson_prb_90} Kivelson, S. A. \& Rokhsar, D. S. Bogoliubov quasiparticles, spinons, and spin-charge decoupling in superconductors. \textit{Phys. Rev. B} \textbf{41}, 11693(R) (1990).

\bibitem{johnson_prl_85} Johnson, M. \& Silsbee, R. H. Interfacial charge-spin coupling: Injection and detection of spin magnetization in metals.\textit{Phys. Rev. Lett.} \textbf{55},  1790  (1985)

\bibitem{bcs} Bardeen, J., Cooper, L. N. \& Schrieffer, J. R. Microscopic Theory of Superconductivity. \textit{Phys. Rev.} \textbf{106}, 162 (1957).

\bibitem{berizinskii} Berezinskii, V. L. New model of the anisotropic phase of superfluid He$_3$. \textit{JETP Lett. \textbf{20}, 287 (1974).}

\bibitem{abrahams_prb_95} Abrahams, E., Balatsky, A., Scalapino, D. J., \& Schrieffer, J. R. Properties of odd-gap superconductors. \textit{Phys. Rev. B \textbf{52}, 1271 (1995)}.

\bibitem{coleman_prb_94} Coleman, P., Miranda, E. \& Tsvelik, A. Odd-frequency pairing in the Kondo lattice. \textit{Phys. Rev. B \textbf{49}, 8955 (1994)}.


\bibitem{ginzburg} Ginzburg, V. L. \textit{Zh. Exsp. Teor. Fiz.} \textbf{31}, 202 (1956).

\bibitem{johnson_apl_94} Johnson, M. Spin coupled resistance observed in ferromagnet-superconductor-ferromagnet
trilayers. \textit{Applied Physics Letters} \textbf{65}, 1460 (1994).

\bibitem{takahashi_prl_99} Takahashi, S., Imamura, H. \& Maekawa, S. Spin Imbalance and Magnetoresistance in Ferromagnet/Superconductor/Ferromagnet Double Tunnel Junctions. \textit{Phys. Rev. Lett.} \textbf{82}, 3911 (1999).

\bibitem{li_prl_13} Li, B., Roschewsky, N. Assaf, B. A., Eich, M., Epstein-Martin, M., Heiman, D., Munzenberg, M., \& Moodera, J. S. Superconducting spin switch with infinite magnetoresistance induced by an internal exchange field. \textit{Phys. Rev. Lett.} \textbf{110}, 097001 (2013)

\bibitem{bulaevskii} Bulaevskii, L. N., Kuzii, and Sobyanin, A. A. \textit{JETP Lett.} \textbf{25}, 290 (1977); Buzdin, A. I., Bulaevskii, L. N., Panyukov, S. V. \textit{JETP Lett.} \textbf{35}, 178 (1982).

\bibitem{buzdin_rmp_05} Buzdin, A. I. Proximity effects in superconductor-ferromagnet heterostructures. \textit{Rev. Mod. Phys.} \textbf{77}, 935 (2005).

\bibitem{bergeret_prl_01} Bergeret, F. S., Volkov, A. F. \& Efetov, K. B. Long-range proximity effects in superconductor-ferromagnet structures. \textit{Phys. Rev. Lett.} \textbf{86} 4096 (2001).

\bibitem{kadigrobov_epl_01} Kadigrobov, A., Shekhter, R. \& Jonson, M. Quantum Spin Fluctuations as a Source of Long-Range Proximity Effects in Diffusive Ferromagnet-Superconductor Structures. \textit{EPL \textbf{54}, 394 (2001).}

\bibitem{bergeret_rmp_05} Bergeret, F.S., Volkov, A. F. Efetov \& K. B. Odd triplet superconductivity and related phenomena in superconductor-ferromagnet structures. \textit{Rev. Mod. Phys.} \textbf{77}, 1321 (2005).

\bibitem{eschrig_prl_03} Eschrig, M., Kopu, J., Cuevas, J. C. \& Sch\"on, G. Theory of Half-Metal/Superconductor Heterostructures. \textit{Phys. Rev. Lett.} \textbf{90}, 137003 (2003). 

\bibitem{keizer_nature_06} Keizer, R. S., Goennenwein, S. T. B., Klapwijk, T. M., Miao, G., Xiao, G. \& Gupta, A. A spin triplet supercurrent through the half-metallic ferromagnet CrO$_2$. \textit{Nature} \textbf{439}, 825 (2006).

\bibitem{eschrig_phystoday} Eschrig, M. Spin-polarized supercurrents for spintronics. \textit{Physics Today} \textbf{64}, 43 (2011)

\bibitem{rashba_soc} E. I. Rashba, \textit{Sov. Phys. Solid State} \textbf{2}, 1109 (1960).

\bibitem{gorkov_prl_01} Gor'kov, L. P. \& Rashba, E. I. Superconducting 2D System with Lifted Spin Degeneracy: Mixed Singlet-Triplet State. \textit{Phys. Rev. Lett.} \textbf{87}, 037004 (2001).

\bibitem{legget_rmp_75} Leggett, A. K. A theoretical description of the new phases of liquid He$_3$. \textit{Rev. Mod. Phys.} \textbf{47}, 331 (1975).

\bibitem{annunziata_prb_12} Annunziata, G., Manske, D., and Linder, J. Proximity effect with noncentrosymmetric superconductors. \textit{Phys. Rev. B} \textbf{86}, 174514 (2012).

\bibitem{bergeret_prb_13} Bergeret, F. S. \& Tokatly, I. V. Spin-orbit coupling as a source of long-range triplet proximity effect in superconductor-ferromagnet hybrid structures 
\textit{Phys. Rev. B} \textbf{89}, 134517 (2014).

\bibitem{dp} D'yakonov, M. I. \& Perel, V. I. Spin Orientation of Electrons Associated with the Interband Absorption of Light in Semiconductors.\textit{Sov. Phys. JETP \textbf{33}, 1053 (1971)}; D'yakonov, M. I. \& Perel, V. I.  Current-induced spin orientation of electrons in semiconductors.\textit{Phys. Lett. A \textbf{35}, 459 (1971)}.

\bibitem{usadel} Usadel, K. D. Generalized Diffusion Equation for Superconducting Alloys. \textit{Phys. Rev. Lett. \textbf{25}, 507 (1970)}.


\bibitem{sau} Lutchyn, R. M., Sau, J. D., and Das Sarma, S. Majorana Fermions and a Topological Phase Transition in Semiconductor-Superconductor Heterostructures. \textit{Phys. Rev. Lett.} \textbf{105}, 077001 (2010).

\bibitem{alicea} Alicea, J. Majorana fermions in a tunable semiconductor device. \textit{Phys. Rev. B} \textbf{81}, 125318 (2010). 


\bibitem{maeno_science_04} Nelson, K. D., Mao, Z. Q., Maeno, Y. \& Liu, Y. Odd-Parity Superconductivity in Sr$_2$RuO$_4$, \textit{Science} \textbf{306}, 1151 (2004).

\bibitem{saxena_nature_00} Saxena, S. S. \etal. Superconductivity at the border of itinerant electron ferromagnetism in UGe$_2$. \textit{Nature (London)} \textbf{406}, 587 (2000).

\bibitem{aoki_nature_01} Aoki, D., Huxley, A., Ressouche, E., Braithwaite, D., Flouquet, J., Brison, J.-P., Lhotel, El. \& Paulsen, C. Coexistence of superconductivity and ferromagnetism in URhGe. \textit{Nature (London)} \textbf{413}, 613 (2001).

\bibitem{asano_prb_05} Asano, Y. Spin current in $p$-wave superconducting rings. \textit{Phys. Rev. B} \textbf{72}, 092508 (2005).

\bibitem{gronsleth_prl_06} Gr{\o}nsleth, M. S., Linder, J., B{\o}rven, J.-M., and Sudb{\o}, A. Interplay between Ferromagnetism and Superconductivity in Tunneling Currents. \textit{Phys. Rev. Lett.} \textbf{97}, 147002 (2006).

\bibitem{brydon1} Brydon, P. M. R., Manske, D., Sigrist, M. Origin and control of spin currents in a magnetic triplet Josephson junction. \textit{J. Phys. Soc. Jpn.} \textbf{77}, 103714 (2008).

\bibitem{brydon2} Brydon, P. M. R., Asano, Y., and Timm, C. Spin Josephson effect with a single superconductor. \textit{Phys. Rev. B} \textbf{83}, 180504(R) (2011).

\bibitem{tanaka_prb_09} Tanaka, Y., Yokoyama, T., Balatsky, A. V., and Nagaosa, N. Theory of topological spin current in noncentrosymmetric superconductors. \textit{Phys. Rev. B} \textbf{79}, 060505(R) (2009).

\bibitem{romeo_prl_13} Romeo, F. \& Citro, R. Cooper Pairs Spintronics in Triplet Spin Valves. \textit{Phys. Rev. Lett.} \textbf{111}, 226801 (2013). 

\bibitem{anwar_arxiv_14} Anwar, M. S., Shin, Y. J., Lee, S. R., Yonezawa, S., Noh, T. W. and Maeno, Y. Prototype hybrid of a ferromagnet and a spin triplet superconductor. \textit{Appl. Phys. Express} \textbf{8}, 015502 (2015).

\bibitem{cc1} Clogston, M. Upper Limit for the Critical Field in Hard Superconductors. Phys. \textit{Phys. Rev. Lett.} \textbf{9}, 266 (1962).

\bibitem{cc2} Chandrasekhar, B. S. A note on the maximum critical field of high-field superconductors. \textit{Appl. Phys. Lett.} \textbf{1}, 7 (1962).


\bibitem{chen_prl_02} Chen, C. D., Kuo, W., Chung, D. S., Shyu, J. H. \& Wu, C. S. Evidence for Suppression of Superconductivity by Spin Imbalance in Co-Al-Co Single-Electron Transistors. \textit{Phys. Rev. Lett.} \textbf{88}, 047004 (2002).

\bibitem{yang_nature_10} Yang, H., Yang, S.-H., Takahashi, S., Maekawa, S., and Parkin, S. S. P. Extremely long quasiparticle spin lifetimes in superconducting aluminium using MgO tunnel spin injectors. \textit{Nature Materials} \textbf{9}, 586 (2010) 

\bibitem{quay_nphys_13} Quay, C. H. L., Chevallier, D., Bena, C., and Aprili, M. Spin Imbalance and Spin-Charge Separation in a Mesoscopic Superconductor. \textit{Nature Physics} \textbf{9}, 84 (2013).

\bibitem{hubler} H{\"u}bler, F., Wolf, M. J., Beckmann, D. \& v. L{\"o}hneysen, H. Long-range spin-polarized quasiparticle transport in mesoscopic Al superconductors with a Zeeman splitting. \textit{Phys. Rev. Lett.} \textbf{109}, 207001 (2012).

\bibitem{yamashita_prb_02} Yamashita, T., Takahashi, S., Imamura, H. \& Maekawa, S. Spin transport and relaxation in superconductors. \textit{Phys. Rev. B \textbf{65}, 172509 (2002)}.

\bibitem{poli_prl_08} Poli, N., Morten, J. P., Urech, M., Brataas, A., Haviland, D. B., \& Korenivski, V. Spin injection and relaxation in a mesoscopic superconductor. \textit{Phys. Rev. Lett. \textbf{100}, 136601 (2008)}.

\bibitem{morten_prb_04} Morten, J. P., Brataas, A. \& Belzig, W. Spin transport in diffusive superconductors. \textit{Phys. Rev. B \textbf{70}, 212508 (2004).}

\bibitem{wakamura_prl_14} Wakamura, T., Hasegawa, N., Ohnishi, K., Niimi, Y., and Otani, Y. Spin Injection into a Superconductor with Strong Spin-Orbit Coupling. \textit{Phys. Rev. Lett.} \textbf{112}, 036602 (2014). 

\bibitem{pdg} De Gennes, P. G. Coupling between ferromagnets through a superconducting layer. \textit{Phys. Lett. \textbf{23}, 10 (1966)}.

\bibitem{hauser} Hauser, J. J. Coupling between ferrimagnetic insulators through a superconducting layer. \textit{Phys. Rev. Lett. \textbf{23}, 374 (1969)}; Deutscher, G. \& Meunier, F. Coupling between ferromagnetic layers through a superconductor. \textit{Phys. Rev. Lett. \textbf{22}, 395 (1969)}.

\bibitem{tagirov_prl_99} Tagirov, L. R. Low-Field Superconducting Spin Switch Based on a Superconductor / Ferromagnet Multilayer. \textit{Phys. Rev. Lett.} \textbf{83}, 2058 (1999).

\bibitem{jiang_prl_95} Jiang, J. S., Davidovic, D., Reich, D. H., and Chien, C. L. Oscillatory Superconducting Transition Temperature in Nb/Gd Multilayers. \textit{Phys. Rev. Lett.} \textbf{74}, 314 (1995).

\bibitem{muhge_prl_96} M{\"u}hge, Th., Garif'yanov, N. N., Goryunov, Yu. V., Khaliullin, G. G., Tagirov, L. R., Westerholt, K., Garifullin, I. A., and Zabel, H. Possible Origin for Oscillatory Superconducting Transition Temperature in Superconductor/Ferromagnet Multilayers. \textit{Phys. Rev. Lett.} \textbf{77}, 1857 (1996).

\bibitem{oh_apl_97} Oh, S., Youm, D. \& Beasley, M. R. A superconductive magnetoresistive memory element using controlled exchange interaction
\textit{Appl. Phys. Lett.} \textbf{71}, 2376 (1997).

\bibitem{fominov_prb_02} Fominov, Ya. V., Chtchelkatchev, N. M., and Golubov, A. A. Nonmonotonic critical temperature in superconductor/ferromagnet bilayers. \textit{Phys. Rev. B} \textbf{66}, 014507 (2002).

\bibitem{zhu_prl_10} Zhu, J., Krivorotov, I. N., Halterman, K., and Valls, O. T. Angular Dependence of the Superconducting Transition Temperature in Ferromagnet-Superconductor-Ferromagnet Trilayers. \textit{Phys. Rev. Lett.} \textbf{105}, 207002 (2010).

\bibitem{singh_arxiv_14} Singh, A., Voltan, S., Lahabi, K., \& Aarts, J. Colossal proximity effect in a superconducting triplet spin valve based on halfmetallic ferromagnetic CrO$_2$. \textit{arXiv:1410.4973}.


\bibitem{leksin_prl_12} Leksin, P. V., Garif'yanov, N. N., Garifullin, I. A., Fominov, Ya. V., Schumann, J., Krupskaya, Y., Kataev, V., Schmidt, O. G., and B{\"u}chner. Evidence for Triplet Superconductivity in a Superconductor-Ferromagnet Spin Valve. \textit{Phys. Rev. Lett.} \textbf{109}, 057005 (2012).

\bibitem{jara_prb_14} Jara, A. A., Safranski, C., Krivorotov, I., Wu, C.-T., Malmi-Kakkada, A. N., Valls, O. T., and Halterman. K. Angular dependence of superconductivity in superconductor/spin-valve heterostructures. \textit{Phys. Rev. B} \textbf{89}, 184502 (2014).

\bibitem{wang_prb_14} Wang, X. L., Di Bernardo, A., Banerjee, N., Wells, A., Bergeret, F. S., Blamire, M. G. \& Robinson, J. W. A. Giant triplet proximity effect in superconducting pseudo spin valves with engineered anisotropy. \textit{Phys. Rev. B} \textbf{89}, 140508(R) (2014).

\bibitem{banerjee_nature_14} Banerjee, N., Smiet, C. B., Ozaeta, A., Bergeret, F. S., Blamire, M. G., and Robinson, J. W. A. Evidence for spin-selectivity of triplet pairs in superconducting spin-valves. \textit{Nature Communications} \textbf{5}, 3048 (2014).

\bibitem{gu_prl_02} Gu, J. Y., You, C.-Y., Jiang, J. S., Pearson, J., Bazaliy, Ya. B., and Bader, S. D. Magnetization-Orientation Dependence of the Superconducting Transition Temperature in the Ferromagnet-Superconductor-Ferromagnet System: CuNi/Nb/CuNi. \textit{Phys. Rev. Lett.} \textbf{89}, 267001 (2002).

\bibitem{moraru_prl_06} Moraru, I. C., Pratt Jr., W. P., and Birge, N. O. Magnetization-Dependent $T_c$ Shift in Ferromagnet/Superconductor/Ferromagnet Trilayers with a Strong Ferromagnet. \textit{Phys. Rev. Lett.} \textbf{96}, 037004 (2006).


\bibitem{potenza_prb_05} Potenza, A. \& Marrows, C. H. Superconductor-ferromagnet CuNi∕Nb∕CuNi  trilayers as superconducting spin-valve core structures. \textit{Phys. Rev. B} \textbf{71}, 180503(R) (2005).


\bibitem{pena_prl_05} Pena, V., Sefrioui, Z., Arias, D., Leon, C., Santamaria, J., Martinez, J. L., te Velthuis, S. G. E., and Hoffmann, A. Giant Magnetoresistance in Ferromagnet/Superconductor Superlattices. \textit{Phys. Rev. Lett.} \textbf{94}, 057002 (2005).

\bibitem{westerholt_prl_05} Westerholt, K., Sprungmann, D., Zabel, H., Brucas, R., Hj{\"o}rvarsson, B., Tikhonov, D. A. \& Garifullin, I. A. Superconducting Spin Valve Effect of a V Layer Coupled to an Antiferromagnetic [Fe/V]  Superlattice. \textit{Phys. Rev. Lett. \textbf{95}, 097003 (2005)}.

\bibitem{miao_prl_08} Miao, G.-X., Ramos, A. V. \& Moodera, J. S. Infinite Magnetoresistance from the Spin Dependent Proximity Effect in Symmetry Driven bcc-Fe/V/Fe  Heteroepitaxial Superconducting Spin Valves. \textit{Phys. Rev. Lett. \textbf{101}, 137001 (2008)}.


\bibitem{eremin_prb_06} Eremin, I., Nogueira, F. S., and Tarento, R.-J. Spin and Charge Josephson effects between non-uniform superconductors with coexisting helimagnetic order. \textit{Phys. Rev. B} \textbf{73}, 054507 (2006).

\bibitem{halasz_prb_11} Halasz, G. B., Blamire, M. G., Robinson, J. W. A. Magnetic coupling-dependent triplet supercurrents in helimagnet / ferromagnet Josephson junctions. \textit{Phys. Rev. B} \textbf{84}, 024517 (2011).

\bibitem{trifunovic_prb_11} Trifunovic, L., Popovic, Z., and Radovic, Z.  Josephson effect and spin-triplet pairing correlations in SF$_1$F$_2$S junctions. \textit{Phys. Rev. B} \textit{84}, 064511 (2011). 

\bibitem{eschrig_nphys_08} Eschrig, M. \& L{\"o}fwander, T. Triplet supercurrents in clean and disordered half-metallic ferromagnets. \textit{Nature Physics} \textbf{4}, 138 (2008).

\bibitem{malshukov_prb_12} Mal'Shukov, A. G. \& Brataas, A. Triplet supercurrent in ferromagnetic Josephson junctions by spin injection 
\textit{Phys. Rev. B} \textbf{86}, 094517 (2012).

\bibitem{anwar_prb_10} Anwar, M. S., Czeschka, F., Hesselberth, M., Porcu, M., and Aarts, J. Long-range supercurrents through half-metallic ferromagnetic CrO$_2$. \textit{Phys. Rev. B} \textbf{82}, 100501(R) (2010).

\bibitem{khaire_prl_10} Khaire, S. T., Khasawneh, M., Pratt Jr., W. P., and Birge. N. O. Observation of Spin-Triplet Superconductivity in Co-Based Josephson Junctions 
\textit{Phys. Rev. Lett.} \textbf{104}, 137002 (2010).

\bibitem{robinson_science_10} Robinson, J. W. A., Witt, J. D. S. \& Blamire, M. G. Controlled injection of spin-triplet supercurrents into a strong ferromagnet. \textit{Science} \textbf{329}, 59 (2010).

\bibitem{sprungmann_prb_10} Sprungmann, D., Westerholt, K., Zabel, H., Weides, M., and Kohlstedt, H. Evidence for triplet superconductivity in Josephson junctions with barriers of the ferromagnetic Heusler alloy Cu$_2$MnAl. \textit{Phys. Rev. B} \textbf{82}, 060505(R) (2010).

\bibitem{houzet_prb_07} Houzet, M. \& Buzdin, A. I. Long range triplet Josephson effect through a ferromagnetic trilayer. \textit{Phys. Rev. B} \textbf{76}, 060504(R) (2007).


\bibitem{waintal_prb_02} Waintal, X. \& Brouwer, P. W. Magnetic exchange interaction induced by a Josephson current. \textit{Phys. Rev. B} \textbf{65}, 054407 (2002). 

\bibitem{zhao_prb_08} Zhao, E. \& Sauls, J. A. Theory of Nonequilibrium Spin Transport and Spin Transfer Torque in Superconducting-Ferromagnetic Nanostructures. \textit{Phys. Rev. B} \textbf{78}, 174511 (2008).

\bibitem{buzdin_prl_09} Konschelle, F. \& Buzdin, A. Magnetic Moment Manipulation by a Josephson Current. \textit{Phys. Rev. Lett.} \textbf{102}, 017001 (2009).

\bibitem{linder_prb_11} Linder, J. \& Yokoyama, T. Supercurrent-induced magnetization dynamics. \textit{Phys. Rev. B} \textbf{83}, 012501 (2011).

\bibitem{teber_prb_10} Teber, S., Holmqvist, C., and Fogelstr{\"o}m. Transport and magnetization dynamics in a superconductor/single-molecule magnet/superconductor junction. \textit{Phys. Rev. B} \textbf{81}, 174503 (2010)

\bibitem{holmqvist_prb_11} Holmqvist, C., Teber, S., and Fogelstr{\"o}m. Nonequilibrium effects in a Josephson junction coupled to a precessing spin. \textit{Phys. Rev. B} \textbf{83}, 104521 (2011).

\bibitem{kulagina_prb_14} Kulagina, I. \& Linder, J. Spin Supercurrent, Magnetization Dynamics, and $\varphi$-State in Spin-Textured Josephson Junctions. \textit{Phys. Rev. B} \textbf{90}, 054504 (2014)

\bibitem{houzet_prl_07} Houzet, M. Ferromagnetic Josephson Junction with Precessing Magnetization. \textit{Phys. Rev. Lett.} \textbf{101}, 057009 (2008).

\bibitem{yokoyama_prb_09} Yokoyama, T. \& Tserkovnyak, Y. Tuning odd triplet superconductivity by spin pumping. \textit{Phys. Rev. B} \textbf{80}, 104416 (2009).

\bibitem{zhu_prl_04} Zhu, J.-X., Nussinov, Z., Shnirman, A. \& Balatsky, A. V. Novel Spin Dynamics in a Josephson Junction. \textit{Phys. Rev. Lett. 92, 107001 (2004)}; Nussinov, Z., Shnirman, A., Arovas, D. P., Balatsky, A. V. \& Zhu, J.-X. Spin and Spin-Wave Dynamics in Josephson Junctions. \textit{Phys. Rev. B \textbf{71}, 214520 (2005)}.

\bibitem{sacramento1} Sacramento, P. D., Fernandes Silva, L. C., Nunes, G. S., Araujo, M. A. N., \& Vieira, V. R. Supercurrent-induced domain wall motion. \textit{Phys. Rev. B} \textbf{83}, 054403 (2011).

\bibitem{sacramento2} Sacramento, P.D. \& Araujo, M.A.N. Spin torque on magnetic domain walls exerted by supercurrents, \textit{European Physical Journal} B \textbf{76}, 251 (2010).

\bibitem{linder_prx_14} Linder, J. and Halterman, K. \textit{Superconducting spintronics with magnetic domain walls}. \textit{Phys. Rev. B} \textbf{90}, 104502 (2014). 

\bibitem{robinson_nature_12} Robinson, J. W. A., Chiodi, F., Halasz, G. B., Egilmez, M., and Blamire, M. G. Supercurrent enhancement in Bloch domain walls. \textit{Scientific Reports} \textbf{2}, 699 (2012).

\bibitem{baker_arxiv_14} Baker, T. E., Richie-Halford, A., and Bill A. Long Range Triplet Josephson Current and 0-$\pi$ Transition in Tunable Domain Walls. \textit{New. J. Phys.} \textbf{16}, 093048 (2014)

\bibitem{blanter_prl_07} Braude, V. \& Blanter, Ya. M. Triplet Josephson Effect with Magnetic Feedback in a Superconductor-Ferromagnet Heterostructure. \textit{Phys. Rev. Lett.} \textbf{100}, 207001 (2008).

\bibitem{kulic_prb_01} Kulic, M. L. \& Kulic, I. M. Possibility of a $\pi$ Josephson junction and switch in superconductors with spiral magnetic order. \textit{Phys. Rev. B} \textbf{63}, 104503 (2001).

\bibitem{pajovic_prb_06} Pajovic, Z., Bozovic, M., Radovic, Z., Cayssol, J. \& Buzdin, A. Josephson coupling through ferromagnetic heterojunctions with noncollinear magnetizations. \textit{Phys. Rev. B} \textbf{74}, 184509 (2006).

\bibitem{buzdin_prb_05} Buzdin, A. \& Koshelev, A. E. Periodic alternating 0- and $\pi$-junction structures as realization of $\phi$-Josephson junctions. \textit{Phys. Rev. B} \textbf{67}, 220504(R) (2003).

\bibitem{asano_prb_06} Asano, Y., Sawa, Y., Tanaka, Y., and Golubov, A. A. Odd-frequency pairs and Josephson current through a strong ferromagnet. \textit{Phys. Rev. B} \textbf{76}, 224525 (2007).

\bibitem{margaris} Margaris, I., Paltoglou, V., and Flytzanis, N. Zero phase difference supercurrent in ferromagnetic Josephson junctions. \textit{J. Phys.: Condens. Matter} \textbf{22} 445701 (2010).

\bibitem{buzdin_prl_08} Buzdin, A. Direct Coupling Between Magnetism and Superconducting Current in the Josephson $\phi_0$  Junction. \textit{Phys. Rev. Lett.} \textbf{101}, 107005 (2008).

\bibitem{sickinger_prl_12} Sickinger, H., Lipman, A., Weides, M., Mints, R. G., Kohlstedt, H., Koelle, D., Kleiner, R., and Goldobin, E. Experimental Evidence of a $\phi$ Josephson Junction. \textit{Phys. Rev. Lett.} \textbf{109}, 107002 (2012).

\bibitem{ryazanov_prl_01} Ryazanov, V. V., Oboznov, V. A., Rusanov, A. Yu, Veretennikov, A. V., Golubov, A. A., and Aarts, J. Coupling of Two Superconductors through a Ferromagnet: Evidence for a $\pi$ Junction. \textit{Phys. Rev. Lett.} \textbf{86}, 2427 (2001).

\bibitem{visani_nphys_12} Visani, C., Sefrioui, Z., Tornos, J., Leon, C., Briatico, J., Bibes, M., Barthelemy, A., Santamaria, J. \& Villegas, E. J. Equal-spin Andreev reflection and long-range coherent transport in high-temperature superconductor/half-metallic ferromagnet junctions. \textit{Nature Physics \textbf{8}, 539 (2012)}


\bibitem{machon_prl_13} Machon, P., Eschrig, M., and Belzig, W. Nonlocal Thermoelectric Effects and Nonlocal Onsager Relations in a Three-Terminal Proximity-Coupled Superconductor-Ferromagnet Device. \textit{Phys. Rev. Lett.} \textbf{110}, 047002 (2013) 

\bibitem{ozaeta_prl_14} Ozaeta, A., Virtanen, P., Bergeret, F. S., and Heikkil{\"a}, T. Predicted Very Large Thermoelectric Effect in Ferromagnet-Superconductor Junctions in the Presence of a Spin-Splitting Magnetic Field. \textit{Phys. Rev. Lett.} \textbf{112}, 057001 (2014).

\bibitem{kawabata_apl_13} Kawabata, S., Ozaeta, A., Vasenko, A. S., Hekking, F. W. J., and Bergeret, F. S. Efficient electron refrigeration using superconductor/spin-filter devices. \textit{Appl. Phys. Lett.} \textbf{103} 032602 (2013)

\bibitem{giazotto_apl_14} Giazotto, F., Robinson, J. W. A., Moodera, J. S. \& Bergeret, F. S. Proposal for a phase-coherent thermoelectric transistor. \textit{Appl. Phys. Lett.} \textbf{105}, 062602 (2014).



\bibitem{moshchalkov1} Lange, M., Van Bael, M. J., Bruynseraede, Y., and Moshchalkov, V. V. Nanoengineered Magnetic-Field-Induced Superconductivity. \textit{Phys. Rev. Lett.} \textbf{90}, 197006 (2003).

\bibitem{moshchalkov2} Gillijns, W., Aladyshkin, A. Yu., Lange, M., Van Bael, M. J., and Moshchalkov, V. V. Domain-Wall Guided Nucleation of Superconductivity in Hybrid Ferromagnet-Superconductor-Ferromagnet Layered Structures. \textit{Phys. Rev. Lett.} \textbf{95}, 227003 (2005).

\bibitem{villegas_sst_11} Villegas, J. E. \& Schuller, I. K. Controllable manipulation of superconductivity using magnetic vortices. \textit{Supercond. Sci. Technol. \textbf{24}, 024004 (2011).}


\bibitem{andreev} Andreev, A. F. Thermal conductivity of the intermediate state of superconductors. \textit{Sov. Phys. JETP} \textbf{19}, 1228 (1964).




}
\end{thebibliography}
\end{document}